# Rural Pension System and Farmers' Participation in Residents' Social Insurance

TAO XU[1]

School of International Economics and Trade

Nanjing University of Finance and Economics

## Abstract

As the ageing population and childlessness are increasing in rural China, social pensions will become the mainstream choice for farmers, and the level of social pensions must be supported by better social insurance. The paper compares the history of rural pension insurance system, outlines the current situation and problems, analyses China Family Panel Studies data and explores the key factors influencing farmers' participation through an empirical approach. The paper shows that residents' social pension insurance is facing problems in the rural areas such as low level of protection and weak management capacity, which have contributed to the under-insured rate, and finds that there is a significant impact on farmers' participation in insurance from personal characteristics factors such as gender, age, health and (family) financial factors such as savings, personal income, intergenerational mobility of funds. And use of the Internet can help farmers enroll in pension insurance. The paper argues for the need to continue to implement the rural revitalisation strategy, with the government as the lead and the market as the support, in a concerted effort to improve the protection and popularity of rural pension insurance.

[1] TAO XU, School of International Economics & Trade, Nanjing University of Finance & Economics, Nanjing, 210046, China.

Contents

# 1. Introduction

In recent years, China's ageing population and the growing number of families with fewer children have evolved into a more serious social problem. Compared to urban areas, the implementation and enforcement of rural retirement and pension policies are more difficult. With the accelerated urbanisation and construction of urban areas, rural young adults are leaving the countryside for cities with better infrastructure in pursuit of better economic conditions, leaving behind elderly people who are retired or have insufficient working capacity. These elderly people have low or no pensions, and rely on the proceeds of their agricultural products and the support of their children to maintain the family retirement.

As a large country with a large rural population, China naturally requires the formation of a sound rural pension system to benefit the majority of farmers. Rural pension system is the key to ensuring the livelihood of the elderly rural population in their old age, and is the main guarantee for their food, clothing and shelter. Compared to the importance of pension insurance in helping the rural elderly, government subsidies or social grants are a drop in the bucket and have little effect. From the New Rural Pension System (NPS) to the Basic Pension System for Urban and Rural Residents, the rural pension system has been undergoing changes since its inception in order to provide maximum protection, as far as possible, for rural residents in their old age. However, reform is always a long process, and there are still many problems with the current rural pension system, such as the low level of protection, which has led to the low income of the rural elderly; the value of frugality and simplicity makes them very cautious about their daily consumption expenditure at relatively low income levels, which indirectly makes it difficult to improve the quality of life of the rural retired elderly. The government's rural pension policy lacks flexibility and practicality, and the current pensions are not sufficient to improve the retirement life of the elderly rural population. In addition, the various surprises and deviations in the implementation or enforcement of the rural pension policy and the participation of the rural elderly people are in need of effective, efficient and reasonable countermeasures to improve the situation.

The earliest academic research on pension insurance in China dates back to the 1980s and focused on the reform of the overall pension system, the development of the pension public utilities and how to raise pension fund (Liqun Wei and Tiejun Li, 1982; Shou Zhang, 1986). In the new century, with the promotion of pension system reform and the introduction of advanced international theories, China's research on pension insurance has moved from the basic "what to do and how to do it" to "how to do it well and what the future should be", and there has been greater progress in qualitative research on related issues. For example, Zilan Liu (2003), based on normative analysis, found that China's pension insurance model must be reformed or reconstructed based on human rights protection, considering land and family, and some socio-economic factors, and China can learn from the case of Brazil with similar national conditions at that time.

As research progressed, more scholars began to turn to quantitative methods that focused on data and transitioned from normative to positive analysis, with more empirical methods. Dasong Deng and Huiyuan Xue (2010) calculated that the establishment of a rural pension insurance system mainly faces problems in terms of local government financing, administration, fund management and farmers' participation. The first three of these reflected the government's administrative capacity, emphasising the question of "what to do and how to do it", while farmers' participation in insurance, the fourth, involved many factors, including the level of government administration as well. Recently, with the implementation of the rural revitalisation strategy, the government has been paying more attention to the old age of the rural population, and apart from the pension gap, other issues at the governmental level have gradually been resolved (Xueliang Liu, 2014). However, there are still many conflicts regarding farmers' participation.

Based on the empirical approach of modern economics, many scholars have studied the factors influencing farmers' participation in insurance and the socio-economic impact of their participation. Zhigang Yuan and Zheng Song (2000) found that regardless of whether pension insurance is pay-as-you-go or fully-funded, an ageing population will incentivise residents to increase their savings to counter the financial risks of retirement, which Lixin He et al. (2008) argued is consistent with the old-age motive for saving, i.e. the life-cycle theory. But, Chongen Bai et al. (2012) explored credit constraints and the motivation to save, arguing that the burden of pension insurance negatively affects current consumption. Using data from the China Family Panel Studies (CFPS), Guangrong Ma and Guangsu Zhou (2014) re-examined the impact of the New Rural Pension System, NPS for short, on household savings and consumption and found that the NPS only reduces the savings rate of the older population rather than others. Although Chen and Zeng (2013) found that NPS reduces the burden of retirement on children and relieves the intergenerational financial pressure of family retirement, it did not have a significant effect on the savings of the younger population. While savings of the younger population may not decline and consumption of the older population may remain meagre, a study by Zong et al. (2015) found that the impact of pension insurance on household investment in risky assets is small and statistically insignificant in rural areas, suggesting that the pressure of retirement of the old age will still hinder the economic contribution of consumption and investment.

Although the new pension insurance system does not immediately improve consumption for the time being, in the long run it may help to increase the purchasing power of the ageing population. Chuanchuan Zhang and Binkai Chen (2014) studied the impact of NPS pension income in the context of intergenerational transfers and found that low levels of NPS do not directly improve social pension capacity. However, NPS pension income can significantly increase the income level of the rural elderly and effectively reduce the incidence of poverty, suggesting that NPS can improve the quality of old age and welfare of the rural elderly and indirectly improve social pension capacity, but, to a certain extent, it will also exacerbate the risk of inequality in the quality of old-age care and retirement services among the rural elderly with different incomes (Chuanchuan Zhang et al., 2016; Ye Zhang et al., 2016). Chuanchuan Zhang et al. (2017) further investigated empirically the impact of NPS implementation on rural people's old-age and retirement expectations and the gender of the birth cohort from the perspective of social retirement expectations and family retirement expectations, and found that the New Rural Pension System has reduced the dependence of the rural population on family supporting, increasing the probability that the rural elderly expect to rely on pensions by about 9% and decreasing the probability that they expect to rely on family supporting by 4%. On this basis, Guangsu Zhou et al. (2020) added that the NPS may not only protect the pensions of the rural population, but also reduces the consumption gap and promotes social distribution equity.

However, not all people have a strong willingness to retire socially, and not all rural people want to participate in the NPS. According to Xiangzhi Kong and Shengwei Tu (2007) and Beihai Tian et al. (2012), the willingness of the rural farming population to participate in social retirement is very weak, as opposed to family retirement, and the rural elderly do have a preference for family retirement. Among them, the need for old-age pension is particularly strong among middle-aged unmarried men in rural areas (Lei Wang, 2015). The demand for rural old-age pension insurance is still expanding, and Houlian Liu (2019) pointed out that a large number of mobile people, the urban-rural immigrants or rural migrant workers in cities, in China, tend to choose self-pension and return to their hometowns, helped by their children; while the demand for social old-age services among rural elderly depends on the demographic structure and the attitude and support of family members (Zhihong Ding, 2014; Juanjuan Sun and Ding Shen, 2017; Zhigang Xu et al., 2018). Jiujie Ma et al. (2021) pointed out that the higher the proportion of elderly people living in the same household, the lower the probability of participation in pension insurance.

From the New Rural Pension System to the Urban and Rural Basic Pension

System, the protection of the rights and interests of the elderly in rural areas has always faced many risks. And, how is it doing now? There is still a lack of systematic analysis of the current problems of the rural pension insurance system, the current situation of participation and the factors influencing it. If one is not insured, then more research on the benefits of insurance is of no practical value, while if one is insured but does not receive the pensions to which one is entitled, then the theoretical benefits of insurance cannot be objectively realised. The paper therefore takes both a qualitative and quantitative approach. Firstly, the paper takes a qualitative approach to the issue of the rural pension insurance system; then, it examines the factors influencing the rate of participation in the pension insurance in a quantitative way; and then, based on the previous analysis, it explores countermeasures and strategies to increase the motivation of farmers to participate in the pension insurance, in order to provide practical suggestions for the implementation of security in old age. The second part of the paper will introduce the development of the rural pension insurance system, outline the current situation and problems. The third part will quantitatively design an experiment, collecting and showing the data. The fourth part will analyse the empirical results of the experiment. The fifth part will summarise the direction of reform based on the results and analysis, and finally conclude the paper.

## 2. History, Problems and Current Situation of Rural Pension System and Farmers' Participation

2.1 History of Rural Pension System

(1) Pension Policies and Provisions in Rural Areas Prior to 2009

In 1949, when New China, the People's Republic of China, was founded, farmers' pensions were all the responsibility of agricultural cooperatives.

After the Reform and Opening Up, the old rural pension insurance policy system became obsolete, and with the rise of the rural economy, such as the household contract responsibility system with remuneration linked to output, farmers' incomes increased, laying the economic foundation for changing the form of contributions to the rural pension system. 1991 saw the introduction of policies such as *the Notice of the Ministry of Civil Affairs on Further Strengthening the Work of Rural Social Pension Insurance* and *the Basic Programme of Rural Social Pension Insurance at the County Level* in 1992, which formally established the rural old-age insurance system, with contributions jointly funded by the government and farmers.

(2) Implementation of New Rural Pension System

Since the new millennium, China's economy has developed rapidly and the income level of Chinese residents is much higher than at the beginning of the Reform and Opening Up period. The rural pension system, which is rooted in the economic foundation of the past, is out of step with the times and the protection it can provide to the majority of rural residents is becoming increasingly meagre and cannot fully meet the needs of the rural elderly. Therefore, in 2009, the State Council issued *the Guidance Opinions of the State Council on Launching Pilot Projects of New Rural Social Pension Insurance*, which formally reformed the rural pension insurance system and introduced a new type of rural pension insurance. The contribution method of the New Rural Pension System is more comprehensive and diversified, consisting of individual contributions, government subsidies, and social grants or collective allowance, and the pension fund is managed at the provincial level.

In Jiangsu Province, for example, the New Rural Pension System provides that those who have already enrolled in rural pension insurance before the age of 60 and those who have received pensions will be eligible for the new rural pension and will receive better protection; those who have not reached the age of 60 and have not received pensions under the old rural insurance will be able to transfer their pensions from their previous accounts to their new rural insurance accounts and will also receive protection under the new rural pension after meeting all the conditions and making contributions in accordance with the standards.

(3) Introduction of Basic Pension System for Urban and Rural Residents

In order to further protect the rights and interests of the elderly, and improve the pension insurance system, the State Council issued *the Opinions on Establishing a Unified Basic Pension Insurance System for Urban and Rural Residents* in 2014, merging the New Rural Pension System and Urban Residents' Social Pension System into the Basic Pension System for Urban and Rural Residents.

The reform of rural pension insurance reflects the process of integration from division to unity, specifically, from the separation of urban workers' pension insurance, farmers' social pension insurance, urban residents' pension insurance and institutions' pension insurance with different levels of protection, to the separation of New Rural Pension System and urban residents' social pension system, and then to the current system of unified management and coordinated levels of protection.

Previously, the pension insurance system was separated between urban and rural areas, and there was a huge difference between the systems, and the problem of convergence, which may be unreasonable and unconscionable, was quite serious. With the promulgation of *the Opinions*, the urban-rural divide was broken down, and *the Opinions* were very much in line with the government's requirements for precise poverty alleviation in the countryside in terms of improving the pension benefits of farmers and integrating the treatment of urban and rural residents. Under the guidance of *the Opinions*, the relevant work has progressed rapidly, and at this time the Basic Pension System for Urban and Rural Residents has effectively alleviated the serious contradictions of the Three Rural Issues①. However, against the backdrop of an increasingly ageing population, there are still many problems with the rural pension insurance system that need to be improved, so its further reform and development is still necessary and essential in order to safeguard democracy and freedom and protect people's rights and interests.

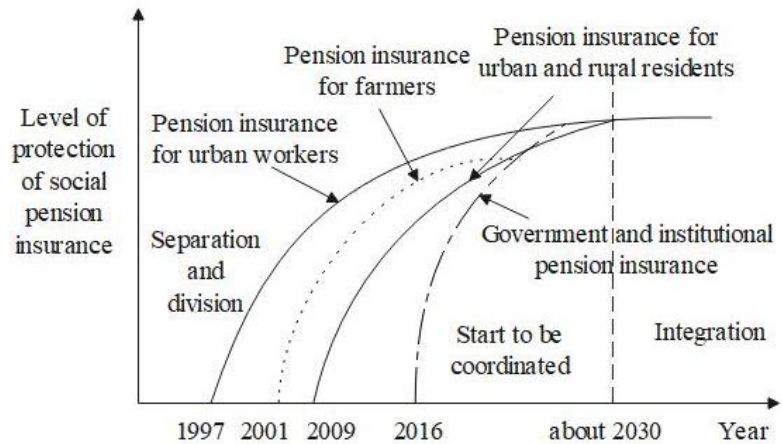

Figure 1. Development of China's integrated social pension insurance system

Notes: The picture is made by the author.

The picture is referenced from the paper Effectiveness of Rural Social Pension System and its Urban-Rural Integration written by Juncheng, Jiang, which was published in *Modern Economic Research* in 2017.

After 2014, the proportion of China's elderly population aged over 65 has remained above 10% every year, growing by 3.4% in seven years, from 139.02 million to 190.64 million. The growth rate of the elderly population is getting higher and higher, and its total proportion to the total population of society is getting bigger and bigger. As China enters a deeply ageing society, there is an urgent need to increase the amount and coverage of pension insurance, and improve the quality of old-age and retirement services.

Table 1. Proportion of elderly population in China

| Year | Number of People Aged 65 and Over (million) | Share of Total Population (%) |
| --- | --- | --- |
| 2014 | 13902 | 10.1 |
| 2015 | 14524 | 10.5 |

---

① Three Rural Issues are issues relating to agriculture, rural areas and farmers, which refers to the improvement of farmers' living conditions, the development of agricultural industry and the progress of society in the rural areas dominated by planting industry. In this century, in the dual society formed by history, cities of China are constantly modernized, and secondary and tertiary industries are developing, but rural progress, agricultural development and farmers' income are lagging behind and at a very low level. Three Rural Issues concretely manifest as the land problems, the basic unit's political power problems, the food security problems, the agricultural policy problems, the farmer individual quality problems and the farmer income problems.

| 2016 | 15003 | 10.8 |
| 2017 | 15813 | 11.4 |
| 2018 | 16658 | 11.9 |
| 2019 | 17603 | 12.6 |
| 2020 | 19064 | 13.5 |

Notes: The data are from *the China Statistical Yearbook*.

2.2  Problems of Rural Pension System

(1)  Overall Protection Level of Rural Pension Is Low and Cannot Cover Needs and Demands of Farmers' Retirement

At present, there is a clear gap between the lower pension levels and what social insurance programs have promised similarly, with some areas even below the subsistence level; most regions have a serious disparity between pension levels and the situation of economic development and consumption. According to Jianhua Yu et al. (2017), farmers tend to choose lower contribution levels, and this reduces the fairness and effectiveness of basic pension insurance protection and security. Contributions to rural pension insurance are mainly paid by farmers themselves, and farmers, caught by their own economic conditions and traditional attitudes, tend to refer to the minimum contribution standard, resulting in low insurance amounts and insufficient future pensions available to farmers to effectively guarantee their lives after retirement.

This situation has improved considerably since the change in the system, with farmers receiving help from village collectives and the government, which has led to more diversified sources of funding for insurance and less pressure to pay contributions. However, even though farmers have paid more in premiums, the protection and security they receive through rural pension insurance is still not enough, and the benefits rural residents receive are clearly inadequate compared to the price level, with inflation reducing the protection in reality even more. After the merger of the New Rural Pension System with the urban residents' social pension insurance system, the national basic pension standard is in the range of 88 to 930 CNY, a protection amount that will not match the future economic situation of the country. Without effective reform, the rural pension insurance paid by these rural residents in the past will certainly not cover their living expenses in the new different situation when they retire in the future.

In addition to farmers' own contributions, pension insurance in rural areas is also subsidised by the government and society, which is of greater benefit to farmers in affluent and wealthy areas where the local government has a generous financial allocation to help share the pressure of contributions. These rural areas, where there are also excellent township enterprises, can further help farmers through transfers to increase rural pension insurance coverage and security. However, this is not possible for rural areas in poorer regions, where the local government is not well-funded and, after its allocating funds to basic housing, health care and education, there are fewer subsidies left for pensions and farmers have to bear the brunt of premium payments. In addition, due to the low level of contributions, the protection provided by the local rural pension insurance is not strong enough, and to a certain extent, the phenomenon of protecting the rich but not the poor has emerged, which can be detrimental to the rights and interests of individuals and, even, to democracy and freedom.

(2) Management of Pension Fund Is Improper and Too Inefficient to Achieve Effective Appreciation

According to policy and research practice, the rate of return on pension fund investments should correspond to the growth rate of the average wage, and the two should match, i.e. the return on investments is slightly higher than the growth in wages. Therefore, Figure 2 of the paper presents the situation after 2012, and since the paper is comparing the rate of return and the growth rate, it is easier to see by the slope, with the average wage corresponding to the left vertical coordinate and the pension insurance balance corresponding to the right vertical coordinate. Figure 2 shows that, in fact, after 2017, pension annuity returns have gradually increased at a lower rate than wage growth. However, with fluctuating inflation, the above simple comparison of rates of return and growth is imperfect.

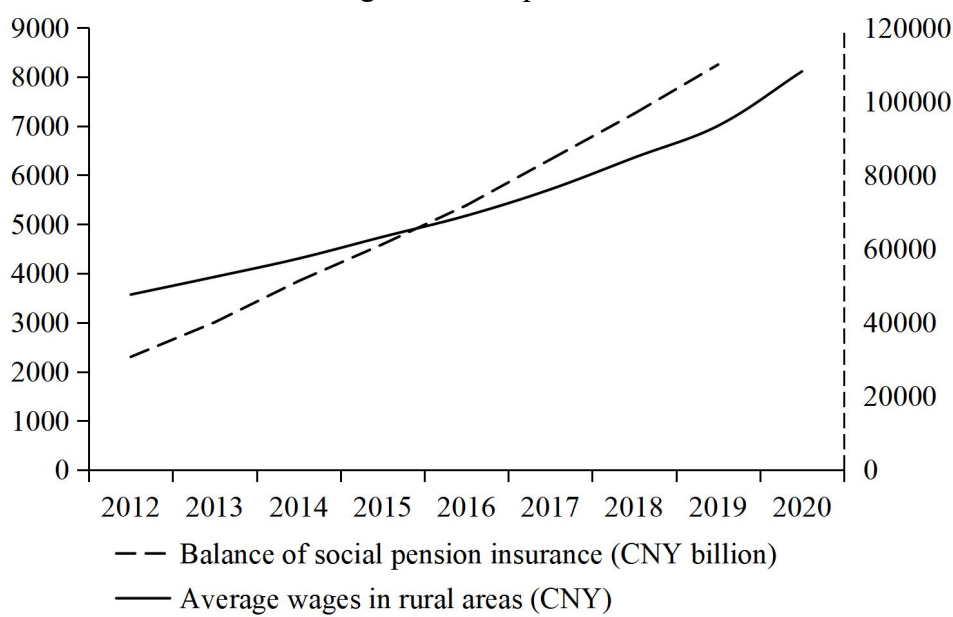

Figure 2. Pension fund appreciation and wage growth in China

Note: The data are from National Bureau of Statistics, PRC.

The dotted line in the graph shows the pension balance, with data corresponding to the right-hand vertical axis; rural wage growth corresponds to the right-hand side.

As shown in Figure 3, taking into account the CPI and GDP deflator, the paper argues that the current returns of pension fund do not meet farmers' needs. The investment of China's pension fund is almost homogeneous, focusing mainly on low-risk investments such as bank deposits and government bonds, and appreciating slowly. In the face of inflation, further improvements in the appreciation of pension fund are urgently needed.

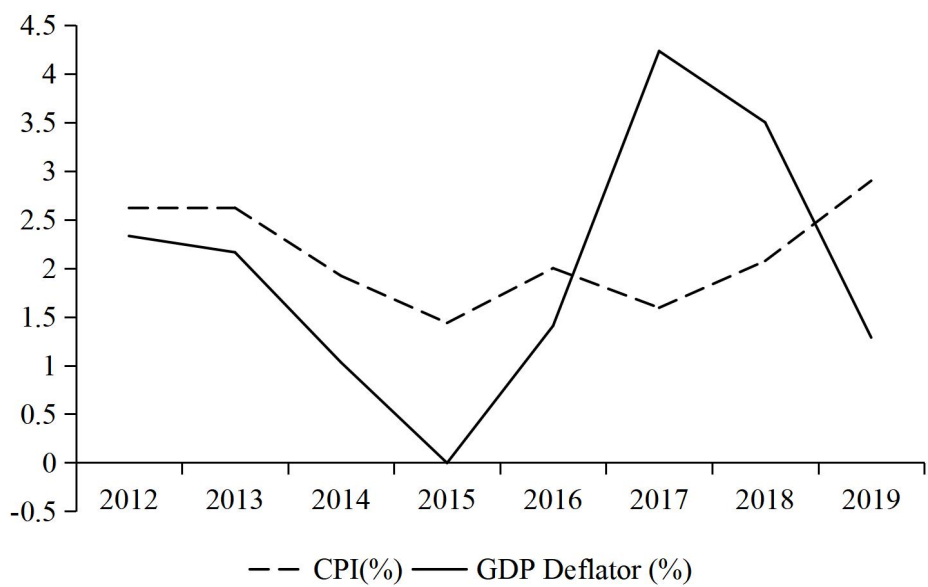

Figure 3. Inflation (CPI and GDP deflator) in China
Notes: The data are from World Bank database.

In addition, there are problems with other management measures of the pension fund. In order to maintain social equity and provide basic protection for the rural elderly, the pension fund, which is made up of urban and rural residents' pension insurance premiums, should have a sound management and monitoring system.

However, in the past, the new rural insurance fund was managed by provincial departments and county-level agencies, and the fund management model for the basic urban and rural residents' pension insurance also provisionally adopted that, but this management model was fragmented and inevitably led to oversights when managing larger amounts, and it lacked strong supervision as it united administrators and supervisors. Such management problems prevented the fund from diversifying and adding further value, and to some extent exacerbate the low returns on rural pension insurance investments. Figure 4 shows the income and expenditure of the pension fund, which may be stable but lack the ability to expand returns. At the same time, although rural pension insurance contributions are mandated to be shared by multiple parties, a large number of village collectives are unable to take on that obligation due to their financial situation or emotional factors, and the authorities are unable to urge all village collectives to help ease the farmers' pressure of payments, making unified management difficult and elasticity insufficient badly.

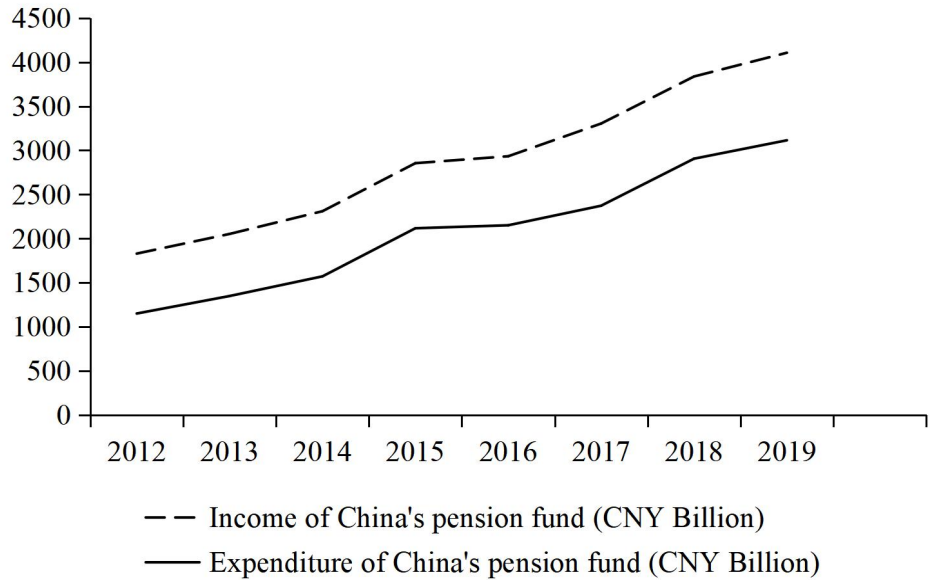

Figure 4. Income and expenditure of China's pension fund

Notes: The data are from National Bureau of Statistics, PRC.

(3) Existing Pension Insurance System Has Produced Uneven Regional Security (Protection) and Poor Urban-Rural Matching (Bridging)

The unevenness and inadequacy of the pension system in terms of location has led to problems such as uneven participation. While rural pension insurance is currently subsidised by the government in terms of payments and contributions, the main part is the responsibility of the farmers themselves yet. Farmers in economically well-off areas may choose to pay higher premiums for better protection, while farmers in poorer areas can often only pay the minimum contribution amount of pension insurance due to economic factors, and the pensions they enjoy after retirement will be much less and lower.

At the same time, the difference and gap in the amount of basic pensions that rural residents can receive in different geographical areas has gradually widened and bigger, with urban and rural residents' pension insurance stipulating contribution levels ranging from 100 CNY per year to 2,000 CNY per year from the lowest to the highest. The majority of farmers in some poor areas tend to pay the minimum standard and receive the lowest level of pensions, with far less economic protection than in developed areas; farmers in economically developed areas are able to pay more premiums and receive more protection, such as Shanghai and Soochow in the southeastern coastal region, where urban and rural residents' pension insurance premiums can reach 2,000 CNY per month, while those in less developed areas in the west can receive is less than 100 CNY per month. Therefore, the construction of the rural pension insurance system is very unbalanced in terms of the combined urban and rural areas and economic location.

Due to the difference in economic level between regions, a large number of farmers from developing areas go to the cities of developed areas to work, and the current system is not flexible in dealing with the situation. It is not only inconvenient for the bridging and connecting (docking or matching) of pension insurance between different administrative regions, but also imperfect between urban and rural areas. The rural pension system now stipulates that it is non-conforming and infeasible to pay premiums in other places, which makes it impossible for many migrant workers to pay in time. And if the farmer pays enterprise worker pension insurance, another form of insurance, in the work site, after returning home they cannot carry on bridging and connecting with rural pension insurance again, making guarantee not in place. Although the policy in some areas allows migrant workers to surrender their urban-related insurance when they return to their hometowns, they still suffer losses in their interests, and their social security rights and interests in urban areas have almost disappeared, which will lead to a worsening of the Three Rural Issues.

(4) Shortage of Professional Personnel Related to Pension Insurance Exists and Relevant Laws and Regulations Are Not Perfect

The specific work of rural pension insurance is currently handled by the county-level and township agencies, while the grassroots work in rural areas is handled by village committees, and most of the matters related to rural pension insurance are coordinated and managed by village cadres. Given the current barriers to entry into the system, a large number of those responsible for rural pension insurance are non-professional and their knowledge of the system is limited to reading the relevant policies, so they are unable to further communicate and convey the policies and answer confusion for farmers because they cannot understand in depth or to the point, and are unable to provide better services to rural residents and further optimise the practical aspects of the pension insurance system. There is such a lack of professionals with rich experience and abundant knowledge in both the optimisation and implementation of the policy, and due to social attitudes and economic factors, there is a scarcity of professionals who can assist in the promotion of the system, which has contributed to the slowdown in the promotion of its coverage and security.

The law on rural pension insurance is limited to *the Insurance Law of the People's Republic of China* and *the Social Insurance Law of the People's Republic of China*, while other relevant regulations are all just several documents, and the laws and regulations are not perfect, not enough, which cannot well establish the legal status of the rural pension insurance system from the institutional level. In addition, there are few laws and regulations to regulate the collection of premiums and the payments of insurance benefits, and there is a risk of negligence in supervision, which could easily give rise to corrupt embezzlement that infringe on the human rights and legal interests of farmers.

2.3 Current Situation of Rural Farmers' Participation

(1) Overall Situation of Farmers' Participation in Rural Pension Insurance

In 2019, the number of participants in China's pension fund totalled 532.66 million, with only 160.32 million actually receiving benefits. Relative to the 190.64 million people aged 65 and above, nearly 16% of the elderly cannot receive deserved pensions. And this elderly population receiving pension insurance benefits is mainly urban people; while the participation rate of rural people, at around 80%, is not low although, but this figure is still insufficient under the system of the People's Republic of China. The specific situation is shown in Figure 5.

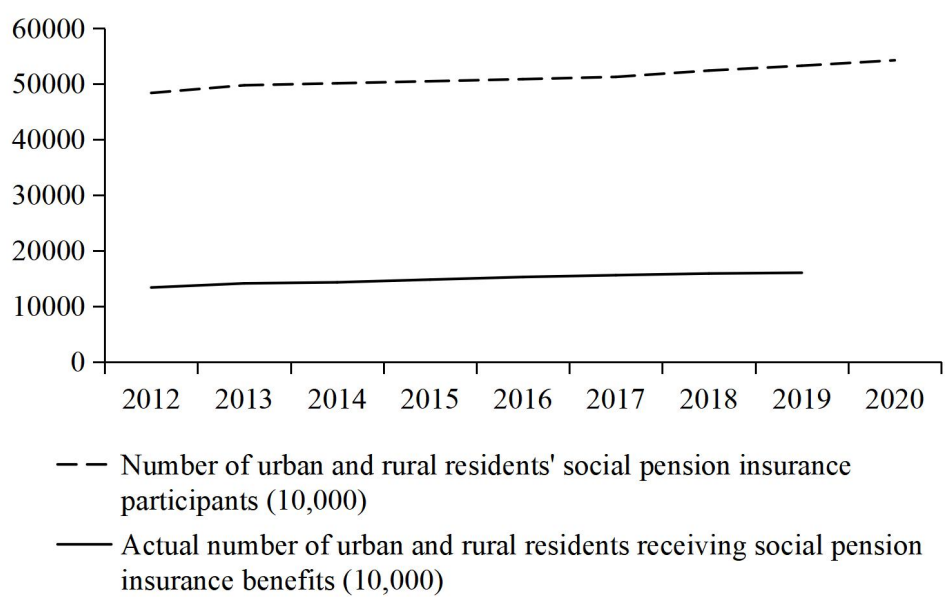

Figure 5. Participation and receipt of pension insurance in China

Notes: The data are from National Bureau of Statistics, PRC.

Farmers' willingness to participate in the rural pension insurance is low, and the government's publicity of the new policy without effective communication is not good, hindering farmers from better understanding the significance of the implementation of the urban and rural residents' pension insurance system, and some farmers are even unaware of its existence. Public announcements and policy explanations about rural pension insurance are basically published on government websites, while some farmers approaching retirement are unable to access the information due to their age and literacy. The traditional concept of the elderly also has a negative impact on the willingness to join the insurance scheme. Most of the elderly in rural areas believe that their old age can be supported by their children, and some of them will look for outsourced jobs after retirement to earn incomes in addition to pensions. The majority of the rural elderly believe in thrift and do not see the need to pay a further sum for social security, and buying commercial insurance is also a waste of money for them.

(2) Farmers' Participation in Rural Pension Insurance under Gender Difference

Previous studies and surveys have shown that the participation rate of male farmers is generally significantly higher than that of female farmers. Adult women who remain in rural areas are influenced by female subordination which should be discarded and tend not to work outside the home, opting more often to become housewives. Traditional attitudes affect women more than men, resulting in rural female residents tending to focus on the family and considering the men in the household as the mainstay. In this case, they may ignore the importance of their own pension security. Considering that the average life expectancy of women is higher than that of men, an inadequate low female participation rate will likely have a direct negative impact on the overall retirement of the future ageing population.

(3) Farmers' Participation in Rural Pension Insurance under Regional Divergence

Figure 6 shows the proportion of pension insurance participants across all provinces in China, and due to space and size constraints, only some examples are presented. As shown, the regions with strong government and developed economy represented by Shandong and Jiangsu account for 14% and 8% of the country's participants respectively, with the southeastern coastal regions accounting for a very large proportion, while the central and western inland regions represented by Qinghai and Gansu have a smaller number.

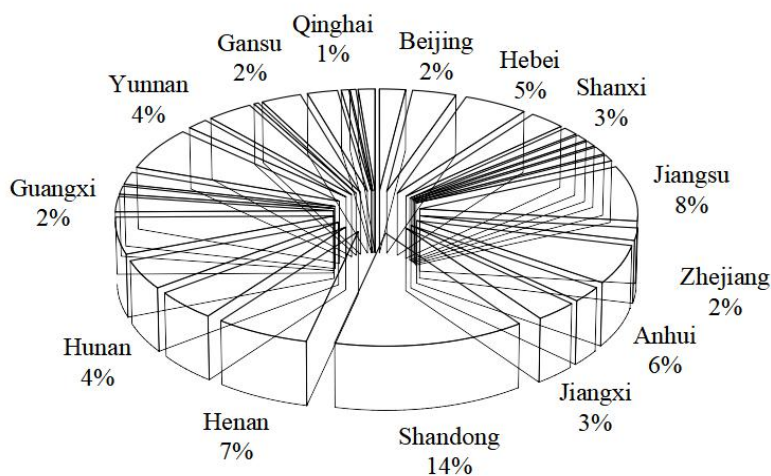

Figure 6. Regional proportion of pension insurance participants (partial illustration)
Notes: The data are from National Bureau of Statistics, PRC.

At present, there is a high degree of migration within China. There are many migrant workers from the countryside to live and work in the cities, which stems from differences in regional economic levels, with rural farmers originally in developing areas preferring to work in cities in developed areas rather than farming locally, getting a new identity as urban workers. As a result, pension insurance participation rates tend to be higher in more developed areas. However, the current system is not flexible enough to properly deal with this phenomenon. It is not possible to effectively match and bridge pension insurance in different regions, nor is it easy to match urban and rural areas as mentioned above. Combining factors such as fund management, overall protection and regional development, farmers' interests suffer a huge loss.

## 3. Materials and Methods

3.1 Samples and Data

The paper will combine empirical methods to analyze the factors influencing participation in rural pension insurance using a sample of rural households from the China Family Panel Studies (CFPS), which has a total sample size of 32,669 but contains a large number of samples that do not meet the requirements of the paper or contain missing values. On the one hand, many urban residents were respondents; on the other hand, many rural residents did not give comprehensive answers to important questions such as basic personal information, household economic situation, information channel preferences and "whether or not to participate in the pension insurance", resulting in some of the sample being incomplete and unusable. Therefore, through a series of data processing, all non-farmers in urban areas or who refused to answer important questions were eliminated from the sample, and the remaining information is collated and considered to be sufficiently randomly distributed, and then quantitative analysis will be conducted to study the factors influencing farmers' participation in rural pension insurance.

The CFPS is a survey conducted and collected by the Institute of Social Science Survey (ISSS) of Peking University, which aims to track and collect data at the individual, household and community levels to reflect changes in China's society, economy, population, education and health, and to provide a basis for academic research and public policy analysis. To date, it covers 25 provinces, municipalities or autonomous regions, with a target sample size of 16,000 households, and includes all members in the sample households.

(1) Dependent Variable

The dependent variable is pension insurance participation in the CFPS data, which is with a "yes=1; no=0" dichotomous characteristic. In the data processing, urban residents have been excluded from the sample and the data below only reflects farmers' participation which will be all in line with the requirements of the experiment.

(2) Independent Variables

The paper selects four major primary variables, which are basic personal characteristics, mainly including individual factors such as gender, age, education and health; family characteristics, mainly including family size, work outside, household savings and intergenerational capital flows, which are financial factors as well; social or economic characteristics (roles of the families), mainly including personal and household income and expenditure; and information channels, which reflect the farmers' preferences in the use of media, mainly including Internet usage, television, print such as newspapers or magazines, and radio or broadcast.

The following specific data are listed as secondary variables as microscopic keys to study the factors influencing rural pension insurance participation.

a. Household savings. Previous studies have concluded that savings and participation in pension insurance show a negative relationship, and that subjects within households are more willing to choose family retirement if savings are higher. To verify the effect of savings on rural pension insurance participation, the paper chose household savings data to capture family savings (Zhigang Yuan and Zheng Song, 2000; Lixin He et al., 2008).

b. Children transfers, i.e. "money given by children", it represents the family supporting, which reflect intra-household financial flows and the intergenerational relationship of pension fund, and are the main support for family retirement, that are used to reflect the impact of the level of family supporting on the willingness to participate in pension insurance (Huashuai Chen and Yi Zeng, 2013; Chuanchuan Zhang and Binkai Chen, 2014).

c. Income. The more income an individual has, the more money he or she can put into pension insurance, but higher income may also cause distortions in the perceptions of the importance of the insurance among the rural elderly, who believe that future savings based on personal income will be sufficient for later life. To investigate the positive or negative correlations, income was included as an variable to test whether an increase in income has a crowding-out effect on rural pension insurance participation (Chuanchuan Zhang et al., 2016; Ye Zhang et al., 2016).

d. Household expenditure, family size, and other relevant factors (Zhihong Ding, 2014; Juanjuan Sun and Ding Shen, 2017; Chuanchuan Zhang et al.). The specific situation of a household is mainly reflected by income and expenditure and size, etc. Larger household size, generally above three, with family members mainly including husband and wife and children. The more children, the larger the household size; while income and expenditure is a representation of a household's economic situation, and the income and expenditure of the household and family members is an additive relationship, with personal income and household expenditure chosen for the paper.

e. Personal characteristics such as age, health condition and marriage inevitably

have an impact on participation in pension insurance; family members cannot accurately predict their later life when they are young and healthy, and it is clearly too late to consider participating in pension insurance after ageing, illness or even widowhood has occurred (Beihai Tian et al., 2012; Xueliang Liu, 2014; Lei Wang, 2015; Liu, 2016).

f. Urban and rural workers, who are out-migration for work (Houlian Liu, 2019).

g. Awareness and use of various information channels, reflecting the potential impact of how policies are disseminated, which is seen in the paper as an opportunity to improve farmers' participation in rural pension insurance.

3.2 Model Specification

The paper will be based on CFPS data, and the empirical analysis will be conducted. The logistic model is used to model the factors influencing participation rates (omitted) as the dependent variable of the model is limited to 0 or 1. The logistic model can be transformed into a linear function. The logit formula is as follows.

$$\text{Logit}(P) = \text{Ln}[P/(1-P)] = \alpha + \sum \beta_i X_i + \varepsilon$$

P is the probability of being participating in pension insurance, 1-P is the probability of not participating in pension insurance. $\alpha$ is the regression constant, $\beta_i$ is the regression coefficient of each dependent variable, $X_i$ is each independent variable, and $\varepsilon$ is the random error term.

3.3 Descriptive Statistics

Descriptive statistics are shown in Table 2. Specific information on number and proportion of people on gender, health status measures, etc. will be presented in the analysis below.

Table 2. Descriptive Statistics

| Primary Variables | Secondary Variables | Mean | Std. Dev. | Assignment |
|---|---|---|---|---|
| | Pension insurance participation | 0.1613833 | 0.3684154 | Yes=1; No=0 |
| Personal characteristics | Gender | 0.6188925 | 0.486055 | Yes=1; No=0 |
| | Age | 46.99023 | 12.77168 | 16 to 81 Years old |
| | Education | 2.711816 | 2.968613 | Illiterate or semiliterate = 0; Primary = 3; ... ; Bachelor = 7; ... Dr = 9 |
| | Health condition | 2.798046 | 1.271444 | Very healthy =1; ... ; Unhealthy = 5 |
| | Marital status | 2.122066 | 0.678497 | Unmarried = 1; ... ; Widowed = 5 |
| Family (financial) characteristics | Family size | 3.686957 | 2.254022 | 1 to 21 People |
| | Out-migration for work | 3.278261 | 1.984868 | Yes=1; No=5 |
| Social and economic characteristics | Household saving | 55664.25 | 135615.6 | 0 to 1000000 CNY |
| | Family Supporting | 1360.522 | 4827.023 | 0 to 35000 CNY |
| | Personal income | 29238.92 | 27100.59 | 0 to 250000 CNY |
| | Household expenditure | 74495.07 | 86415.57 | 1916 to 918572 CNY |
| | Government subsidy | 3.104348 | 2.001632 | Yes=1; No=0 |
| Channels of information | Internet | 0.3342939 | 0.4724243 | Yes=1; No=0 |
| | Television | 3.443804 | 1.395341 | 1 to 5 |
| | Print | 1.56196 | 1.071731 | 1 to 5 |
| | Broadcast | 1.801153 | 1.282699 | 1 to 5 |

Notes: The data are from CFPS.

## 4. Empirical Results

4.1 Baseline Regression

(1) Impact of Personal Characteristics on Rural Pension Insurance Participation

Table 3's (1) shows that gender and age have a significant effect on farmers' participation in rural pension insurance, the coefficient of gender is -0.69 and the coefficient of age is 0.29, the former is significant at the 1% level and the latter is at the 10% level. In terms of the age of the sample, the coefficient is positive and the dominance ratio is greater than 1, indicating that there is a positive relationship between increasing age and rural pension insurance participation; in terms of the gender of the sample, the ratio of men to women in the sample is 31:19 according to National Bureau of Statistics and *China Statistical Yearbook*, and Table 3 shows that the dominance ratio is 0.5, indicating that men are 2.1 times more likely to participate than women. This is because women are more inclined to be supported by family due to family conditions, social status and ideology, while rural men are more involved in socio-economic activities and more likely to accept social retirement and pensions.

(2) Impact of Family Characteristics on Rural Pension Insurance Participation

Table 3's (2) shows that the significance of gender in individual characteristics increases, while health status becomes significant, the coefficient of health status is -0.75, the dominance ratio is less than 1, indicating that the healthier the person, the less inclined to participate in pension insurance. Among household characteristics, household savings and money given by children have a significant effect on participation in it, with a negative coefficient for household savings and a dominance

ratio of less than 1 as shown in Table 3, and a positive coefficient for money given by children with a dominance ratio greater than 1. Of these, money given by children is significant at the 5% level. The increase in household savings has a negative relationship with participation of rural pension insurance, which is contrary to the conclusions of some previous studies; the increase in intergenerational mobility of money has a positive relationship with rural pension insurance participation.

(3) Impact of Social and Economic Characteristics on Rural Pension Insurance Participation

In (3) of Table 3, it is shown that the significance of gender in individual characteristics returns to 10% from 1%, and the effect of health is more significant than in (2); household savings in household characteristics becomes insignificant. As to socio-economic (roles of the families) terms, personal income has a significant effect on participation in rural pension insurance with a positive coefficient. The more personal income one has, the more willing one is to participate in rural pension insurance. In the experimental sample, there is no crowding-out effect in this respect. But government subsidies may backfire by making farmers content with the status quo, resulting in their non-participation due to lack of attention and cooperation.

(4) Impact of Information Channels on Rural Pension Insurance Participation

In (4) of Table 3, the impact of the other three types of characteristics is shown to be similar to (3). Among the information channels, Internet use are more significant, with the coefficient of Internet use being 1.97, significant at the 5% level. The more and more frequently the Internet is used, the more willing one is to participate in rural pension insurance. The opposite is true for the print media, so replacing the print media with the Internet can increase the motivation to participate.

Table 3. Baseline results (coefficients and dominance ratios)

| Participation | (1) | | (2) | | (3) | | (4) | |
|---|---|---|---|---|---|---|---|---|
| | Logit | Logistic | Logit | Logistic | Logit | Logistic | Logit | Logistic |
| Gender | -0.6848994* (0.427928) | 0.504141 (0.21574) | -1.636683** (0.699788) | 0.1946245 (0.136196) | -2.313211*** (0.817899) | 0.098943 (0.080925) | -2.818562*** (0.990610) | 0.0596917 (0.05913) |
| Age | 0.2868373*** (0.034512) | 1.332207 (0.04598) | 0.2896993*** (0.056665) | 1.336026 (0.075706) | 0.3507434*** (0.070621) | 1.420123 (0.100291) | 0.4805621*** (0.107605) | 1.616983 (0.17310) |
| Health condition | -0.0876177 (0.156356) | 0.916111 (0.14324) | -0.7488076** (0.302121) | 0.4729301 (0.142882) | -0.8419436*** (0.354615) | 0.4308723 (0.152794) | -1.106801*** (0.414893) | 0.330615 (0.13717) |
| Household saving | | | -0.0000153* | 0.9999847 | -0.000022* | 0.999978 | -0.0000274* | 0.9999726 |

| | | | | | | |
|---|---|---|---|---|---|---|
| Family supporting | (0.000009) | (0.000009) | (0.000012) | (0.000012) | (0.000014) | (0.00001) |
| | 0.000181** | 1.000181 | 0.0002107** | 1.000211 | 0.0002121** | 1.000212 |
| | (0.000083) | (0.000083) | (0.000092) | (0.000092) | (0.000107) | (0.0001) |
| Personal income | | | 0.0000406* | 1.000041 | 0.0000512** | 1.000051 |
| | | | (0.000023) | (0.000023) | (0.000026) | (0.000026) |
| Government subsidy | | | -0.3559225* | 0.7005269 | -0.3638889* | 0.6949684 |
| | | | (0.1895039) | (0.1327526) | (0.212293) | (0.147537) |
| Internet | | | | | 1.966165** | 7.143232 |
| | | | | | (1.116913) | (7.978372) |
| Print | | | | | -0.8491257* | 0.4277888 |
| | | | | | (0.496871) | (0.212556) |
| Other variables | Controlled | | Controlled | | Controlled | Controlled |
| R^2 | 0.4331 | | 0.4597 | | 0.5143 | 0.5765 |
| Sample size | 1125 | | 1125 | | 1125 | 1125 |
| F test | - | | - | | - | - |

Note: *** P<0.01, significant at 1% level; ** P<0.05, significant at 5% level; * P<0.1, significant at 10% level. The table shows coefficients (above) and standard error values (below in parentheses). Same below.

### 4.2 Robustness Test

After the analysis of the baseline regression results, the paper will conduct robustness tests by replacing variables, changing the model and regressing sub-samples. Firstly, the dependent variable "pension insurance participation" is replaced with specific "urban and rural residents' pension insurance participation" in the baseline model using the variable substitution method. In the experimental questionnaire, the former is more general, while the latter is more accurate focusing on the new specific form of pension insurance. As shown in Table 4, after the replacement of the dependent variable, the regression results do not change substantially from the baseline regression results.

Table 4. Logit regression results after replacing the dependent variable (coefficients)

| Participation | Logit | | | |
|---|---|---|---|---|
| | (1) | (2) | (3) | (4) |
| Gender | -0.1597828 | -0.8518126 | -1.050536* | -1.427115** |

| | (0.3407) | (0.5578) | (0.5977) | (0.6945) |
|---|---|---|---|---|
| Age | 0.2000043*** | 0.2194646*** | 0.2390844*** | 0.3623849*** |
| | (0.0241) | (0.0420) | (0.0474) | (0.0746) |
| Health condition | 0.013985 | -0.4728237** | -0.4159124* | -0.6147628** |
| | (0.1237) | (0.2317) | (0.2384) | (0.2980) |
| Marital status | 0.479852** | 0.6837387* | 0.6905425* | 0.41598 |
| | (0.2073) | (0.3917) | (0.3980) | (0.3812) |
| Household saving | | -9.66E-06* | -8.84E-06 | -0.0000103 |
| | | (0.0000) | (0.0000) | (0.0000) |
| Family supporting | | 0.0001327** | 0.0001386** | 0.000103 |
| | | (0.0001) | (0.0001) | (0.0001) |
| Internet | | | | 2.770814*** |
| | | | | (0.8878) |
| Television | | | | 0.437093*** |
| | | | | (0.2593) |
| Print | | | | -8.64E-01*** |
| | | | | (0.4981) |
| Other variables | Controlled | Controlled | Controlled | Controlled |
| $R^2$ | 0.2963 | 0.3490 | 0.3587 | 0.4779 |
| Sample size | 1125 | 1125 | 1125 | 1125 |
| F test | - | - | - | - |

Secondly, the model was changed to replace the logit model with the probit model and regression analysis is performed and the results are shown in Table 5. On the basis of changing the model, the regression results are not significantly different from the logit baseline regression results.

Table 5. Probit regression results after replacing the model (coefficients)

| Participation | Probit | | | |
|---|---|---|---|---|
| | (1) | (2) | (3) | (4) |
| Gender | -0.4229165* | -1.000731*** | -1.351079*** | -1.561256*** |
| | (0.2277) | (0.3882) | (0.4530) | (0.5113) |
| Age | 0.1519666*** | 0.1649125*** | 0.1958077*** | 0.2555477*** |
| | (0.0167) | (0.0296) | (0.0358) | (0.0501) |
| Health condition | -0.0784396 | -0.4546504*** | -0.4914312** | -0.6132504*** |
| | (0.0858) | (0.1757) | (0.2002) | (0.2238) |
| Household saving | | -9.18E-06* | -0.0000121* | -0.0000137* |
| | | (0.0000) | (0.0000) | (0.0000) |
| Family supporting | | 0.0001004** | 0.0001182** | 0.0001141* |
| | | (0.0000) | (0.0001) | (0.0001) |
| Personal income | | | 0.0000233* | 0.0000296** |
| | | | (0.0000) | (0.0000) |
| Government subsidy | | | -0.1905935* | -0.1692078 |
| | | | (0.1005) | (0.1071) |
| Internet | | | | 1.010286* |
| | | | | (0.5671) |
| Print | | | | -0.4618611* |
| | | | | (0.2732) |
| Other variables | Controlled | Controlled | Controlled | Controlled |
| R^2 | 0.4335 | 0.4638 | 0.5157 | 0.5669 |
| Sample size | 1125 | 1125 | 1125 | 1125 |
| F test | - | - | - | - |

It is worth adding that in the second part, the paper provides a qualitative analysis of the impact of regional development on rural pension insurance participation. Therefore, the paper then conducts sub-sample regressions to exclude economically developed regions represented by Shanghai, Jiangsu, Zhejiang and Guangdong. As shown in Table 6, the sub-sample regression results and the baseline regression results are uniform and harmonised, showing good robustness.

Table 6. Logit regression results for subsamples (coefficients)

| Participation | Logit | | | |
|---|---|---|---|---|
| | (1) | (2) | (3) | (4) |
| Gender | -0.6635869 | -1.570413 | -2.552736** | -5.634278** |
| | (0.5937) | (1.0471) | (1.2786) | (2.6539) |
| Age | 0.3091472*** | 0.2539457*** | 0.378551*** | 0.7960079*** |
| | (0.0485) | (0.0742) | (0.1137) | (0.2969) |
| Health condition | -0.1711539 | -0.8659256* | -1.389775** | -2.92211*** |
| | (0.2107) | (0.4836) | (0.6377) | (1.2150) |
| Household saving | | -0.000026 | -0.0000456* | -0.0000393 |
| | | (0.0000) | (0.0000) | (0.0000) |
| Family supporting | | 0.0001791* | 0.000286 | 0.0004734 |
| | | (0.0001) | (0.0001) | (0.0002) |
| Government subsidy | | | -0.6961759** | -0.2897785** |
| | | | (0.3242) | (0.3515) |
| Internet | | | | 4.499434* |
| | | | | (2.8300) |
| Television | | | | -1.063598* |
| | | | | (0.6306) |
| Print | | | | -2.482376** |
| | | | | (1.3071) |
| Other variables | Controlled | Controlled | Controlled | Controlled |
| R^2 | 0.4397 | 0.3593 | 0.4726 | 0.6882 |
| Sample size | 252 | 252 | 252 | 252 |
| F test | - | - | - | - |

## 4.3 Explanation and Interpretation of Results

At the personal level, age and gender have a major impact on participation in rural pension insurance. The higher the age, the more willing one is to participate, as the pressure of retirement problems becomes heavier with ageing, which in turn gives individuals a psychological incentive to participate; the gender issue has been mentioned in many recent studies and will not be repeated here (Xiansheng Lei, 2020; Peng Zhan, 2020; Hongbo Jia and Ge Gao, 2020).

At the family level, the level of household savings and intergenerational financial flows have a major impact on participation in rural pension insurance. There is a negative relationship between household savings and willingness to participate, with the higher the savings, the lower the willingness to participate. The higher the intergenerational mobility, i.e. the more money children give to their parents, the more willing they are to participate.

In terms of socio-economic terms (roles), the negative impact of income is greater. The greater the individual's income, the greater the confidence in the family retirement, and the less inclined they are to choose to enroll in rural pension insurance. Although the household's income and expenditure did not pass the significance test due to the sample, following the line of analysis of Yating Ren et al. (2016), the paper augurs that the more household expenditure in rural areas, the higher the willingness to participate in insurance will likely be, which is contrary to Ren's view.

The impact of the above three aspects on farmers' participation in pension insurance has been analysed in some depth in existing studies. The importance of the Internet in terms of social and information channels is also of great concern. According to information sorted out by Tao Xu et al. (2021), an increasing proportion of middle-aged and elderly people are now using the Internet, and they use it to try to obtain more information when they are under pressure to retire. The Internet, empowered by new technologies such as big data and cloud computing, has the ability to break information imbalance and create an efficient information matching mechanism. Under the system of PRC, the relevant public authorities or departments can collect these records through the aforementioned advanced technologies to help rural residents solve their old-age problems, providing better services.

The use of the Internet is a wonderful opportunity to reform the rural pension system and improve farmers' participation in the scheme. As they age and their physical fitness declines, the elderly rural population will no longer be able to engage in outdoor activities at high frequency and intensity, and the development of the mobile Internet provides an excellent means of entertainment. Now that China has achieved total poverty eradication, the willingness of rural residents with some savings to deploy mobile Internet devices for communication and connection, based on their personal income and the help of their children, is on the rise. Against the backdrop of technological innovation, the function of mobile phones and the Internet for the middle-aged and elderly is gradually being upgraded from simple communication and information gathering to conversation, two-way transmission of information and relatively sophisticated entertainment, permeating all aspects of life. This has created good conditions for strengthening the promotion of pension insurance through "Internet+government" and fostering the enthusiasm of rural residents to participate in pension insurance.

By investing in new media information about pension insurance and increasing relevant tweets on platforms such as WeChat and Alipay, which are commonly used by rural residents as well, rural residents can directly or indirectly become more aware of information about rural pension insurance through the social act of retweeting in various forms, such as online forwarding or offline conversation, which can enhance their trust in the pension insurance system while eliminating the information gap and improve their willingness to participate in the insurance in a subtle way for the more information available, the better it will help in decision making. Through digital and online pension insurance service providers, relevant policies, news and fraud prevention measures can be released in a variety of forms on related platforms, so that rural residents who are accustomed to using the Internet can experience the safety of the financial environment and the convenience of services in an entertaining way, thus increasing their motivation to participate in insurance.

At the same time, the participation of rural residents cannot only be improved proactively by the above-mentioned path, but also through big data to accurately identify and collect information. In business, enterprises use the Internet to break the information mismatch and realise the reverse transmission mechanism of demand guiding supply in terms of commodities; while in the rural pension insurance system, the government also completes a top-level design that is more in line with the national conditions based on the information and data transmitted in the reverse direction, starting from the actual needs of rural residents, so as to attract them to join the insurance with the high quality and level of the rural pension insurance system itself.

## 5. Countermeasures for Development of Rural Pension System

5.1 Improving the System

(1) Government Leads to Improve and Increase Protection of Rural Pension Insurance from Multiple Perspectives

Although the empirical analysis does not highlight the impact of the level of pension insurance coverage and security, it must be taken seriously as a fundamental solution to the pension problem.

Firstly, the basic pension level must be increased in line with the realities of the situation. The basic pension of urban and rural residents' pension insurance has been greatly improved compared to the minimum monthly pension of 55 CNY in rural areas in the past, but the future socio-economic development after the 14th Five-Year Plan will tend to be of high quality. According to Huaizhong Mu et al. (2020) and Fangxiao Hu et al. (2021), the pension insurance system suffers from inadequate levels and imbalances in funding and benefits, while Wei Zheng and Youji Lu (2021) also make similar observations, arguing that pension insurance contributions and their level of protection are incompatible with the demographic structure, national wealth distribution and income levels, and that the current situation cannot adapt to high-quality socio-economic development.

The current basic pension standard is already insufficient in some areas and will not be able to effectively protect farmers' retirement in the future. In addition, a large number of the elderly rural population prefer to go out to work when they are young, and most of them work in industries that require strenuous physical labour, such as construction and manufacturing, which has caused some of the elderly rural people to fall ill and suffer from many illnesses in their old age. However, now, the insurance benefits provided by rural pension insurance, both in the past and at present, are unable to meet the medical and health care needs of farmers. In summary, the government needs to increase the amount standard of the basic pension (Hua Yang, 2015; Ruiqin Gao and Jingzhong Ye, 2017; Jieyu Liu and Jiaqing Yu, 2019).

Secondly, subsidies and benefits should be increased appropriately. In addition to the fixed subsidies stipulated in the policy, local governments can, depending on their own economic situation, allocate a further sum of money to subsidise farmers or use it to build a comprehensive elderly care system. For poor farmers, they should carry forward the spirit of the state's precise poverty eradication efforts to increase subsidies and provide more benefits to rural elderly people. The government can also encourage local enterprises or collective economic organisations to pay part of the premiums for farmers to give back to society, or build a series of infrastructures in townships such as nursing homes and universities for the elderly that can help improve the lives of farmers in their old age, building a reputation as a caring enterprise, further promoting local economic development and breaking the mould for the current situation in the countryside.

In addition, Hongqian Qi and Yan Yang (2020) pointed out that it is also necessary to achieve the preservation and appreciation of the value of pension insurance funds. In order to ensure that urban and rural residents have something to fall back on in their old age, the government has imposed restrictions on the investment of pension accounts, which to a certain extent prevents the elderly from having their pension money harvested, which is a good thing, but also makes pensions slow to increase in value and even risk devaluation if inflation occurs. With the current blossoming of "Internet+finance" investment projects and the further implementation of tougher regulation of Internet finance, China will build a more complete "Internet+finance" investment model in the future and the government could also introduce policies to allow the elderly in rural areas to invest their pensions in capital-protected financial products on officially regulated Internet platforms, helping farmers to reap more dividends and improving the protection and appreciation of their pension insurance in many ways.

(2) Government Helps to Reform and Optimise Fund Management Methods

After the merger of rural pension insurance and urban residents' pension insurance into urban and rural residents' pension insurance, the government has implemented a series of measures to improve the management system of urban and rural residents' pension insurance, such as crediting the urban and rural residents' pension insurance fund to the special account of the social security fund and using different books to separately account for income and expenditure to guarantee the independence and authenticity of the data, thus avoiding financial falsification to a certain extent; gradually converging the urban and rural residents' basic pension insurance fund from the previous situation of management at the municipal and township levels to management at the provincial level, and promoting the operation of pension fund on a unified national platform to facilitate investment. These initiatives have optimised the investment channels and regulatory model for pensions, and are an improvement over the rural pension insurance of the past.

However, as the first line of protection for the interests of the rural elderly and an important support for the government in addressing the Three Rural Issues, the pension insurance still needs further optimisation, which should not just be treated lightly. The government can try to gradually liberalise some of the restrictions on pension investment in the capital market and make the investment means more diversified, such as buying government bonds with higher yields and stronger liquidity, or starting with long-term investments in the stock market, which often show an increasing trend in the rate of return when holding shares for a long period of time, and avoiding the volatility of short-term investments, so that farmers will not lose their money overnight, so that pension investment channels are no longer limited to bank deposits and government bonds, which to a certain extent helps to preserve and enhance the value of pensions.

According to Chunli Zhang (2016) and Yi Zeng et al. (2019), as pensions serve as the material dependence of farmers for the rest of their lives, the government's supervision of fund operations must necessarily be strengthened. Firstly, the government should increase the supervision of pension fund, strengthen the control of each process including fund raising, allocation, investment, management and disbursement, and confirm the flow of each fund, and the finance and audit departments, as the frontline departments of supervision, must take up their responsibilities. Secondly, technological innovations should be made to the way the fund is supervised, using information technology such as big data and blockchain to track the various processes of pension management and update the information and status of pension recipients in real time, so as to prevent the emergence of illegal situations such as fraudulent claims, duplicate claims, and claims after death, and to make the pension management system more stringent. Finally, information transparency should be achieved. In addition to internal supervision by the government, rural residents should also be involved in the supervision of pension fund, so as to achieve social supervision. As farmers' own interests are at stake, they are bound to be concerned about the management and operation of pensions, which in effect strengthens the publicity effect of the basic pension insurance for urban and rural residents. The rural pension insurance system should be established as a trinity of laws and regulations, government departments and the people to further optimise fund management practices.

(3) Government and Social Organisations Coordinate Bridging Work to Reduce Regional Disparities

However, due to various factors such as the economy, working rural residents are highly mobile and may return to their villages at any time, suspending the payments of urban workers' pension insurance, and delaying the payments of urban and rural residents' pension insurance premiums while working outside the cities, which ultimately provides insufficient protection.

To solve such problems, Qin Zhu (2020) pointed out that the government could introduce more official and formal policies to adjust the interface and bridging work between urban and rural residents' pension insurance and urban workers' pension insurance, for example, stipulating that farmers should enroll in urban workers' pension insurance if they can earn a stable income in enterprises while working in cities, and enroll in urban and rural residents' basic pension insurance when they are unemployed, so that the two can be dovetailed and farmers can pay their insurance premiums without any interruption or default. At the same time, the government should also speed up the construction of a platform, a nationwide unified social insurance service platform, and strive to co-ordinate the systems under regional differences, so as to narrow the gap between the pension levels of economically developed and developing regions.

### (4) Government and Institutions Keep Refining Legal Mechanisms and Work on Nurturing More Professionals

The basic pension insurance for urban and rural residents came into effect in 2014, and since then more and better laws and regulations have emerged to safeguard the retirement of rural residents, but they are still not comprehensive enough. Although *the Social Insurance Law of the People's Republic of China* and *the Opinions of the State Council on the Establishment of a Unified Basic Pension Insurance System for Urban and Rural Residents* protect the implementation of the basic pension insurance system for urban and rural residents from the general direction, when it comes down to concrete implementation at the grassroots level, these rules and regulations are often not implemented due to their lack of perfection, so more detailed documents and files are needed. The government should continue to strengthen the system, invite experts from home and abroad in economics, finance, sociology and law on rural pension insurance to discuss the construction of the system and establish a nationally unified public service platform for social insurance, rather than simply assigning the task to local governments to unfold autonomously in accordance stiffly with regulations and policies, which would neither comply with the laws nor truly give convenience to farmers because in practise the action may be full of surprises and deviations in detail that will cause confusion.

With regard to the lack of professionals in the implementation of the system, the government can establish a special programme with universities to train a group of professionals who know the system and have insights, and select them to go to areas where the implementation of urban and rural residents' pension insurance is not smooth, so that these professionals can show their talents in their respective fields, solve various problems related to pension insurance in each place, and explain the policy to rural residents and do practical work well.

### 5.2 Improving the Participation

### (1) Government Tries to be Innovative in Approach to Publicity and Carefully Helps Farmers Get Accustomed to Social Retirement

The Internet has been widely accepted by rural pension insurance participants and is proving to be an effective way to promote the scheme. The government's past approach to promoting rural pension insurance has been proven ineffective by the market, and the promotion of urban and rural pension insurance today requires a more effective approach of publicity, and it is not enough to solely post announcements, documents and files on government websites that farmers rarely visit or not access at all. The government's promotion of rural pension insurance might as well make use of these platforms and invite some self-publishers, who may be celebrities and KOL, to explain the policies of urban and rural residents' pension insurance to farmers, which not only has an effect on the promotion, but also helps establish a friendly, sociable and meaningful cyberspace community, responding to the country's call to regulate the reasonable use of Internet.

At the same time, for rural residents with limited literacy and unfamiliar with the Internet, local township offices or village committees can help villages update their bulletin boards and build large electronic screens to scroll through the policies on the pension insurance system and assign people to explain the benefits of purchasing it to farmers under boards.

Personal factors such as gender, age and health, and family financial factors such as income, savings and intergenerational support have proven to be very important influences on farmers' participation in insurance. With thousands of years of smallholder economic foundations in rural China, history has naturally shaped the concept of land retirement, with support by children. In this context, a large number of rural residents do not sufficiently trust government policies and regard buying insurance designed by the government as "irrational consumption", believing that they have their own income and insisting on the support of their children as the primary choice of retirement, supplemented by their own money. Rural women, in particular, are reluctant to take out social pension insurance due to social traditions, economic status and ideological influences. For these rural residents, the government should play the role of grassroots cadres in the townships and assign cadres who have a good reputation among the village people and experience in their work to explain the purpose and objectives of the government's rural pension insurance and urban and rural residents' pension insurance, etc., and to convey the universality of the government's relevant policies, that contributions are a "rational investment" for their future old age after retirement rather than an "irrational consumption". Young rural residents who are healthy and have a stable source of income should not disregard the financial risk of future illness treatment just because they are temporarily unwell, nor should women be resistant to joining the insurance because of the gender difference between men and women. In addition, misunderstandings about the policy must be avoided with reasonable attitudes and detailed explanations to address the problem of misinformation.

(2) Government Implements Rural Revitalisation Strategy and Adjust Farmers' Income from the Root

Empirical evidence has found that excessive savings may lead to a decrease in farmers' willingness to participate, while improved income and financial support from their children increase the incentive to participate. It is therefore important to adjust farmers' relative regional incomes, take into account their traditional attitudes and foster their trust in social pension insurance.

"All systemic problems are ultimately economic problems." In 2016, the State Council issued *the 13th Five-Year Plan for Poverty Eradication*, which aims to achieve full poverty eradication by 2020, and the rural poverty alleviation work has benefited the largest group of farmers. In order to further consolidate the effect of poverty eradication and solve the Three Rural Issues, the state still needs to make more efforts to improve the structural income level of farmers. On the one hand, green agriculture or rural tourism can be developed to revitalise the traditional farming industry; on the other hand, policies can be introduced to attract talents to set up businesses in rural areas, absorbing surplus rural labour and facilitating local employment for rural residents to drive economic development. All in all, the government can introduce policies to consolidate the results of poverty alleviation, and after raising farmers' income, use consumption to curb excessive savings, then not only will the participation rate of urban and rural residents' pension insurance be raised, but farmers' living well-being level will also be improved, and some existing contradictions can certainly be alleviated.

(3) Government Introduces Additional Commercial Insurance as Appropriate to Collaboratively Better Rural Old-Age and Retirement Services

It is worth adding that commercial insurance can also be put to good use. Social insurance is introduced by the government to meet the basic livelihood protection of residents. From rural social pension insurance to urban and rural residents' pension insurance, the government has made many efforts to protect the interests of farmers, and social pension insurance has covered most of the rural residents, but the protection of social insurance is weak after all, and its basic pension can not cover the living expenses of the elderly often, at this time if the commercial pension insurance intervention, the The rural residents will be better protected in their old age with the intervention of commercial pension insurance.

Commercial insurance is available in a range of types, premiums, and insurance options that farmers can choose from, and is tailored to provide better service and more protection. Commercial insurance companies have professional risk management teams and investment and financial management channels, and pension fund can be better managed by professionals to preserve and increase their value, providing rural residents with a higher level of retirement protection. By guiding commercial pension insurance to the countryside, it will not only provide more protection for farmers, but also reduce the financial burden of paying pensions. However, commercial pension insurance in the market today is a mixed bag, with some commercial insurance companies pursuing profits at the expense of customer needs. Therefore, the government must be strict in guiding commercial pension insurance into the rural market and introduce it appropriately, without overdoing it.

# 6. Discussion and Conclusions

The development of the rural pension insurance system is inextricably linked to the changes in the economic situation. Both the new rural pension insurance system and the basic pension insurance system for urban and rural residents have emerged from corresponding situation. Under the 14th Five-Year Plan, the social and economic development will be of higher quality and the rural areas will face a new situation. The rural pension system must be constantly improved to keep pace with the times.

The paper summarises the three major historical stages of the rural pension insurance system, outlines the current situation and problems, and finds that rural pension insurance is facing problems such as low levels of protection, weak management capacity and too little number of participants. In order to address the issue of participation, the CFPS tracking survey data have been selected and logistic and logit model has been used to conduct an empirical study on farmers' participation in rural pension insurance. Specifically, the better the health condition, the more reluctant farmers are to join the scheme, and men are more willing to join it than women; personal income and intergenerational support show a positive effect, while savings have a negative one. At the same time, the paper innovatively finds that the Internet can help farmers better participate in pension insurance, in a way that breaks the information mismatch and helps the government to increase the participation rate in rural areas.

At the time when the population is ageing and the siphoning effect of large cities is increasing, the rural population is losing its young and middle-aged counterparts, and the elderly population is becoming more and more "left behind or semi-left behind". It is urgent to improve the rural pension insurance system and promote the construction of a social pension insurance system for urban and rural residents. The paper suggests that it is necessary to implement the strategy of rural revitalisation, with the government as the leader and the market as the support, from various perspectives such as legal, economic and moral, around the personal characteristics of farmers, financial situation and information channels, using the Internet, to improve the protection power and popularity of rural pension insurance, and provide multi-level pension protection services.